# My 10-Day App Crash Course in China: An Autoethnography

YUE FU, Information School, University of Washington, US

This paper presents an autoethnography of my recent trip to China, during which I engaged in using various apps and discovered the cultural and social norms embedded in everyday mobile app use. Navigating between Western and Chinese cultures, my experience was simultaneously exhilarating, embarrassing, and bewildering. Through this autoethnography, I aim to illustrate how I adjusted to Chinese technological norms, usage patterns, and interactions during my initial stay, and to offer observations on the technosocial differences related to smartphone apps in both cultures. Using descriptions and summative analyses, I identified four meaningful themes: 1) smartphones as the backbone for modern living in China, 2) smartphone attachment, 3) the superapps, and 4) the intricate web of Chinese technosocial norms governing everyday usage. Taken together, these findings highlight how cultural and societal differences shape app design and user experiences, consequently influencing how travelers and expatriates adapt to an increasingly digitalized world.

## 1 Introduction

Recently, I traveled to China for a working/vacation trip. Having grown up there before moving to the U.S. over a decade ago, this was my first extended stay, lasting over a month. During my initial time there, I found myself struggling to adjust to the Chinese technosocial culture. My use of technology was clumsy and often met with amusement or ridicule from strangers and my elderly parents. As a researcher in Human-Computer Interaction (HCI), I found this experience both fascinating and bewildering. Conducting this autoethnography is the means to understand how a cross-cultural traveler like myself navigates the paradox of using technologies and apps that are seemingly familiar but, in essence, different in two distinct cultural contexts. My cross-culture background affords me to observe, analyze, and reflect from both the Chinese and American perspectives, through which I aim to shed light on how the technosocial norms differ between Chinese and American cultures, and how the process of learning to use local apps can serve as a lens into its culture context.

In this paper, I present an autoethnography [12], sharing my experiences during the first ten days of the trip, providing insights into my adjustment to everyday technologies, apps, and my observation of technosocial norms of people's everyday technology use contrasted with usage in the U.S. During this period, I collected extensive fieldnotes [34], captured screenshots of my app usage, and documented observations of myself and others through photos and my interpretation. I maintained daily journals summarizing and reflecting on these observations, expanding them into thick descriptions [19]. Following established autoethnographic methods and examples [16, 25, 34], I conducted a summative analysis to extract overarching themes, resulting in four meaningful themes: 1) smartphones as the backbone for modern living in China, 2) smartphone attachment, 3) the superapps, and 4) the intricate web of Chinese technosocial norms governing everyday usage.

By sharing these experiences and reflections, I aim to contribute to a deeper understanding of how technosocial differences between cultures influence people's lifestyles, norms, and interactions related to smartphone usage. My experience shows that learning to navigate local apps is not just a technical skill, but also a critical step toward understanding and integrating into a community's cultural fabric.

Author's Contact Information: Yue Fu, chrisfu@uw.edu, Information School, University of Washington, Seattle, Washington, US.

## 2 Related Work

### 2.1 Autoethnography in HCI

Autoethnography is a qualitative research method focusing on personal experience to understand cultural phenomena [16]. It has gained increasing traction in HCI and researchers have used autoethnographic approaches to understand diverse technology experiences, often challenging traditional HCI's emphasis on objective, third-party knowledge [25]. This method, drawing on anthropological traditions [10], positions the researcher as both participant and observer [31], allowing for an exploration of the interplay between personal experience and broader cultural contexts [34]. HCI studies have employed autoethnographic approaches to explore the effects of location-based services on a bus driver's work [32], examine learning experiences through music [39], and probe the use of wearable devices to increase awareness of time [20]. Lucero's study of living without a mobile phone demonstrates the potential of autoethnography for deeply personal inquiries into technology's role in everyday life [25]. This body of work acknowledges that the researcher's experiences are inherently shaped by their cultural background and biases but also aims to maintain scholarly rigor by embracing reflectivity [33]. The present autoethnography follows this tradition, focusing on the intricacies of navigating an unfamiliar technosocial culture and how it shapes everyday technology use and cultural adaptation.

### 2.2 Mobile Apps and Cross-Cultural Adjustment

Previous research highlights the important role of mobile apps in facilitating expatriates' adjustment to different cultures, particularly in language learning and understanding local customs [24, 27]. Active engagement with smartphone applications can enable smoother cultural adaptation, while resistance or ineffective use may hinder this process [38]. Studies have shown that cultural distance can impair the positive effects of mobile device usage on knowledge-sharing intentions among expatriates [22]. Additionally, researchers have introduced the concept of "digital distance"—the differences in technology use across cultures—which should be considered alongside traditional cultural distance when investigating individuals' experiences and adjustment processes [38].

Despite these works, the ways in which expatriates and travelers adapt to technosocial practices and how mobile apps influence their lives remain "under-researched and under-recognized," particularly regarding people's day-to-day experience [38]. This autoethnography seeks to provide a personal account of the struggles and reflections associated with adjusting during an initial ten-day stay in China. By sharing my experience, this study aims to shed light on the complexities of cross-cultural adaptation facilitated or impeded by mobile technologies.

## 3 Case

In this section, I provide an overview of my initial ten-day stay, on which this autoethnography is based. I will describe my technological background, daily routine, and situate my observations and technology use within the environment of a rapidly digitizing China.

I lived in China until my early twenties before moving to the United States over a decade ago. At the time of my departure, mobile technology was just beginning to emerge, with mobile apps being rudimentary and often mirroring their web portal counterparts. I acquired my first smartphone a year before leaving China, with fewer than ten apps installed—mostly for communication, weather updates, and news.

During my years in the United States, both China and the USA experienced a mobile technology boom that dramatically transformed every aspect of people's lives. China's embrace of mobile technology accelerated at an unprecedented pace. For example, in the realm of mobile payments, China skipped the adoption of the credit cards phase, which is common for Western countries, transitioning directly from a predominantly cash-based society to one where mobile app payments are ubiquitous. By 2019, China had approximately 1.05 billion mobile payment users [36]. The two most popular payment apps, *WeChat Pay* (微信支付) and *Alipay* (支付宝), had penetration rates of 92.4% and 72.1%, respectively [36]. Both my personal experience and business analyses [18, 26] indicate that China has embraced mobile technology much faster than the United States, which only began wider adoption, particularly in mobile payments, during the COVID-19 pandemic in 2020.

Recently, I traveled back to China and stayed at my parents' apartment in a large eastern Chinese city. The apartment is located in the city's central residential district, close to all forms of public transportation—subway, buses, shared bikes, etc. My daily routine typically begins around 5:00 am when I wake up. I start the day with meditation and prayer for 30 minutes. Following that, I work on various academic projects, including paper revisions, new manuscripts, and preparation for a course I will teach in early 2025. After breakfast with my parents, I usually take a break to browse social media and watch short videos using Chinese apps like *Bilibili* [2] (哔哩哔哩, similar to YouTube) and *Douyin* [3] (抖音, the Chinese version of TikTok, owned by ByteDance). I then return to work until 11:00 am when we have lunch.

After lunch, I often take a nap. Before drifting off, I often find myself lying in bed, scrolling through social media and watching more short videos again. The afternoons are my free time. I often go out for various activities such as going to the gym, practicing piano, shopping, playing badminton, or simply exploring the city on a shared bike (I subscribed to a monthly shared bike service on *Meituan* "美团" for 13.8 Chinese yuan, approximately $1.91). During this time, I use my phone extensively for searching for deals and activities, navigation, listening to podcasts, watching videos (especially at the gym), consuming news, and communicating with merchants, friends, and family members. Around 5:00 to 5:30 pm, I return home to have dinner with my parents or meet friends and relatives for dining out. If I dine out, I use apps to find great restaurant deals, purchase coupons, and check out using the same app, which is often much cheaper than paying the bill on the spot.

I usually start another work session around 6:30 pm, reflecting on the day's activities, expanding and updating my fieldnotes, organizing screenshots, and conducting other academic-related tasks. I go to bed around 9:00 pm.

Although I had traveled to China before, those visits were usually much shorter and primarily spent visiting friends and family. I didn't have much personal time or the need to explore independently, so I didn't find it necessary to download various Chinese apps during the short stay—most activities were organized by friends and relatives who also generously covered expenses. This trip, however, was different. For the first time, I had an extended period in China and aimed to explore different aspects of living there—shopping, signing up for a gym, renting shared bikes, buying groceries, and more. In essence, I wanted to experience life as a resident and insider rather than being pampered by others.

The motivation for this autoethnography emerged during my first two days back in China. I found myself struggling to use various Chinese apps, and even the process of signing up felt foreign, requiring me to learn new technosocial norms and interactions. Within those initial two days, my phone battery died twice (I suspect Chinese apps are more battery-intensive than their American counterparts). On the second day, I discovered I had no phone and data service while heading to a haircut appointment, unaware that I had already exhausted my service plan. Suddenly, I was lost in my childhood city without service, did not know where I was, and was unable to call for help. Eventually, I managed to restore my service, but subsequently the hairdresser refused to accept cash, leaving me embarrassed and unable to settle the bill on the spot. These bewildering experiences during the first two days with technology left me amusingly frustrated, particularly as I myself am an HCI researcher. They highlighted the profound technosocial differences between Chinese and Western cultures and motivated me to undertake this autoethnography.

## 4   Method

In line with established autoethnographic methods [16, 25, 34], I grounded the study in descriptions of daily experiences, combining them with reflective analysis to extract meaningful themes, providing an in-depth examination of how I navigate and adapt to technosocial norms in Chinese cultural contexts. The following outlines the data collection and analysis process.

**Daily Journal:** I have maintained a habit of journaling daily since middle school. During my trip to China, I continued this practice, dedicating 30-45 minutes each morning to reflect on the previous day's experiences. This journaling was more extensive than my usual routine, where I typically spend 5-10 minutes. In my China trip journals, I focused on documenting not only the events of the day but also my reflections on them, particularly related to technology usage and my interpretation about the technosocial environment at large.

**Observation Fieldnotes and Notes Interpretation:** Throughout the day, especially during my exploratory outings, I took extensive fieldnotes using my iPhone's native apps, Reminders and Notes. These brief notes, often consisting of keywords and in-situ reactions, served as triggers for later reflection and interpretation. They allowed me to capture real-time observations about my interactions with technology and the technosocial environment in China. I also photographed contexts and scenes relevant to my observations. In the evenings, I revisited my notes, expanding them into detailed writing blocks.

**Serendipitous conversation with parents and relatives:** I did multiple informal conversations with my parents and relatives about on their technology use. These interactions were spontaneous, often occurring during meals, walks, or shopping trips. While I occasionally prompted discussions on technology use, many times these topics emerged organically or initiated by my parents and relatives. After each conversation, I recorded immediate reflections and insights on my Iphone, which were then expanded and contextualized during my morning journaling sessions. This allowed me to compare my own observations with those of others who were native to China and had experienced these technosocial norms for much longer.

**Screenshots:** I captured various screenshots during my interactions with different mobile apps to document my digital traces. These included login screens, ads displayed within apps, gamification features, and interaction flows such as purchasing deals, renting and returning shared bikes, and navigating public transportation. I also periodically took screenshots of my phone's system data, for example, on battery consumption, which appeared to be much higher compared to my usage in the United States. At the end of each day, I uploaded these screenshots to an online collaborative tool, Figma [4], where I organized them according to specific interaction sessions and themes.

**Data Analysis:** I began the data analysis process by organizing all collected materials. This included arranging my daily journals in chronological order, compiling all fieldnotes and related summaries into a separate document, and categorizing screenshots by interaction sessions within Figma. I then conducted a thorough read-through of the data, taking note of recurring patterns and observations. A summative analysis [15] was performed to identify and categorize emerging themes. Then I systematically grouped related data and extracting specific cases that illustrated the initial themes. As I progressed, I refined and merged some themes to form a more coherent narrative.

I shared my initial draft with colleagues and friends who travel frequently between China and the U.S. Their feedback helped clarify and validate certain aspects of the technosocial norms and struggles I observed. I also translated the draft into Chinese using ChatGPT [29] and shared it with my parents, whose insights facilitated my interpretation of how my experiences as a returnee differed from those of native users in adapting to technology use in China.

In the following sections, I present my findings on how I experienced and adjusted to the technosocial culture of China during my initial ten days.

## 5 My 10-Day App Crash Course in China

### 5.1 Smartphones as the Backbone for Modern Living in China

In modern-day China, owning and knowing how to use a smartphone is a necessity. Navigating daily life, particularly in urban environments, without a smartphone is nearly impossible. Smartphones, especially *superApps* (also called "everything-apps," which provide a multitude of services; see section 5.3 for details), have become deeply integrated into everyday activities, transforming the way people interact with services and one another. From paying for meals via QR codes to using mobile apps for everything from transportation to medical service, the smartphone is essential for survival in a big Chinese city.

On the second day of my trip, I experienced firsthand how integral smartphones are to life in China. I booked a haircut deal using the *Meituan* mobile app (美团, a comprehensive superapp for lifestyle services) and planned my route using the *Gaode Map* (高德地图, similar to Google Maps*)*. After paying for my subway ticket via *WeChat Pay*, everything seemed in order. However, as soon as I stepped off the subway and tried to use the map, I realized my phone had run out of data. A message popped up saying I needed to pay, but without internet service, I couldn't call anyone or use the mobile network operator's app. To make matters worse, I began to panic since I realized I had no cash to pay for my subway return trip since my parents and friends suggested that I not bring cash often. I searched for a Starbucks, the only place I could think of that could offer free Wi-Fi, but after failing to find one, I walked aimlessly until I stumbled into a shopping mall. There, I finally found a place to connect to Wi-Fi, top up my data, and get back on track. As I made my way to the salon, I couldn't help but reflect on how dependent I was on my phone. Without it, I was essentially lost—unable to navigate the city, make purchases, or communicate.

Despite this, I didn't fully learn my lesson. On day three, I decided to treat my mom to a nice dinner followed by shopping for clothes. I carried emergency cash and a credit card, only to discover that the restaurant only accepted mobile payments, and my cash wasn't enough. My mother ended up paying, which was embarrassing since I had intended to cover the bill. This issue repeated itself when we went shopping, with the only exception being at *UNIQLO*, which is the only store that accepted my U.S. credit card. This experience highlighted just how dependent modern Chinese life has become on smartphones. Without them, people can easily find themselves stranded or unable to complete basic tasks.

The prevalence of *superapps* accelerates the dependency on smartphones. At a cafe near my parents' apartment, orders were placed through two superapps: *WeChat Mini Programs* (微信小程序) and *Meituan* (美团). There was no physical menu. The waitstaff often had to explain to customers how to use these apps, and at times, even took over customers' phones to complete the orders. Ordering digitally was more advantageous, offering discounts and coupons. At the cafe, I saved up to 40% using online coupons, and leaving positive reviews could earn rewards like free desserts. I also noticed that roughly 50% of the staff's interactions with customers were related to using technology—explaining how to order, how to save money by buying various bundles online, redeeming coupons, and helping customers upload ratings. Customers, in turn, were often engrossed in discussions about how to navigate the various digital offers and discounts within themselves.

These apps are designed to encourage constant engagement, integrating technology into every part of daily life. Even when I did not intend to spend much time on my phone, micro-interactions (often linked with numerous ads, discounts, coupons, etc.) increased my time spent on my smartphone. In fact, my iPhone screen time increased by 43% during my week in China. The deep integration of these technologies into everyday routines makes owning and using a smartphone indispensable for life in Chinese cities. As my mother succinctly put it: "*As long as you have a phone and a bank card associated with it, you can travel across China without any problem.*"

## 5.2 Smartphone Attachment: The Normalization of "Phubbing"

The pervasive integration of smartphones into daily life has led to a deepening attachment between people and their devices. This attachment has raised concerns globally, with research highlighting the phenomenon of smartphone addiction, the manipulation of user behavior through app design, and the potential negative impacts of excessive phone use. Public discourse, especially in China, has also addressed these issues, with strict regulations on minors' gaming and screen time. However, based on my observations, the attachment to smartphones in everyday life is ubiquitous, permeating various social settings in China.

The following vignettes illustrate the extent of smartphone attachment that I observed during my time in China:

(1) **Subway Journey:** On my way to a haircut, I took the subway during off-peak hours. Most passengers were sitting, with a few standing. I noticed everyone around me was glued to their phones. The only exception was a monk in a yellow robe, standing calmly, gazing out the window.

(2) **Dinner at a Restaurant**: While dining at a large restaurant with my mother, I observed a couple sitting nearby. Throughout their meal, one partner remained engrossed in their phone, only briefly lifting their eyes for minimal conversation. The other partner barely spoke and ate quietly. Behind them, another couple had their phones on the table, staring at their screens with earphones on, exhibiting what I call the "phone smirk"—a half-laugh or smile triggered by the content on their devices.

(3) **Couples on the Street:** I observed several couples walking hand in hand on the streets. While they held hands, their other hands were firmly attached to their phones, each looking in opposite directions, absorbed in their own digital worlds.

(4) **Family TV Time:** At home, my parents were watching a TV show in the living room with the volume turned up, yet both were simultaneously using their smartphones. My father played a Chinese chess game with sound on, while my mother scrolled through short videos. The room echoed with the sounds of three different devices—two smartphones and the television.

(5) **Study Session at the Library:** During a visit to the public library, I noticed that out of 22 individuals in the study room, 10 were using their smartphones. Among the phone users, four exhibited the "phone smirk."

(6) **Tea Shop Experience:** At a famous local tea shop, I noticed that all five salespeople were focused on their phones, oblivious to my presence. I waited for about 20 seconds before any of them noticed me.

These vignettes, though just a sample of my observations, illustrate how entrenched smartphone use has become in daily life.

During a casual dinner conversation, I asked my parents about their smartphone habits. My mother, an accountant, reported using her phone for about six hours during a working day, primarily for reading and consuming news. My father, a retired manager, admitted he didn't track his phone usage, but my mother quickly remarked, "*He must use it longer than me.*" Both acknowledged that they used their phones out of boredom and noted the cognitive toll it took on them. My mother explained that smartphone use had made it difficult to engage in deep thought, with most of the information they consumed being forgettable. They both expressed difficulty in distinguishing high-quality information from low-quality content and acknowledged the pervasive lure of instant gratification.

Interestingly, I also observed that people in China did not seem embarrassed when caught using their phones or engaging in phubbing. One evening, I asked my mother if she felt awkward using her phone when someone else walked into the room for a conversation or assistance. She explained that she used to feel self-conscious, but as phone use became ubiquitous and other people seemed undisturbed, it no longer felt problematic for her. My observations in cafes, restaurants, parks, gyms, and my own family validate her statement. Over time, *phubbing* (phone snubbing) had normalized, and the behavior was integrated into the cultural fabric, changing both individual and collective perceptions of acceptable social conduct.

### 5.3 The Prevalence and Impact of Superapps

*Superapps* have become a cornerstone of daily life in China, fundamentally transforming how people interact with technology. These "everything-apps" provide a multitude of services—including payment processing, messaging, finance management, local information, transportation services, and more—supporting both personal and commercial activities [5]. The two most widely recognized superapps are *WeChat* (微信) and *Alipay* (支付宝), each offering more than 20 functionalities. However, during my initial stay in China, I discovered that these are not only exceptions: most apps I used embraced the superapp design concept, integrating functions beyond their primary purposes.

For example, *Gaode Map* (高德地图), originally designed as a navigation tool similar to Google Maps, now offers services and information on news, weather, e-commerce, ride-hailing, tourism, marathon sign-ups, house rentals, and more. This can be bewildering for those accustomed to Western apps, which tend to be more focused and single-purpose. The superapp design typically includes multiple navigation tabs and hidden menus, as the front page cannot accommodate all features. Each tab or function has its own deep information architecture, akin to a standalone app. In the case of WeChat, users can access millions of mini-programs developed by third parties—similar to plugins in an integrated development environment (IDE)—without transitioning to other apps [1]. I found the superapp concept comparable to visiting a multi-story mega mall, common in Asia, where nearly every service and shop is available within one complex, divided by floors and districts. Customers navigate within the mall to find what they need, and there are interactions between different stores, such as cross-mall coupons and rewards. In contrast, Western apps can be likened to individual stores scattered across a sprawling suburban city, where each provides distinct services without shared systems, requiring customers to travel separately to each location.

Not only do these superapps offer multiple functions, but their interface design is often busy and complex. Accustomed to Western minimalist interface design, I initially found superapps overwhelming, with numerous buttons, text blocks, floating icons, pop-up ads, and more. It was also challenging to understand each sub-function, and I unconsciously ignored deeper branches of the information architecture. However, when I discussed this with my relatives and friends, they did not share the same sentiment. Most found superapps to be extremely helpful, providing supportive functions they needed. This divergence in app design paradigms made me question the generalizability of canonical Western usability heuristics, such as Nielsen's usability heuristics [28], which are widely adopted in interface design in the U.S.

Over time, I began to appreciate the major benefit of superapps: convenience. However, I also perceived multiple drawbacks, which I summarize below:

(1) **Extended Use and Unconscious Time Wasting**: As mentioned in the previous section, ordering food in a restaurant using a superapp can take a considerable amount of time. The process involves searching for deals, calculating rewards and coupons, and navigating pop-up ads and promotional games. Compared to ordering with a physical menu, it often takes much longer. Some of the superapps are layered with *"navigation and financial fog"* [8]: app interfaces use confusing elements, unclear language, financial rewards, and lengthy and complex interaction flows to create navigation "*fog*" to keep users on the apps. For instance, while ordering at a cafe using *Meituan* (美团), I encountered various coupons, rewards, and discounts that were not clearly explained. My bill decreased from 42 yuan to 26.8 yuan after applying the initial discounts, but the process was confusing. An additional discount was offered to deduct 5 yuan for a payment of 0.01 yuan, and finally, I got a deduction of 1.99 yuan, which did not mathematically add up when applied. After several screens and prompts—some leading to in-app games and others offering vague rewards like "*Meituan coins*"—I spent over three minutes just to order a cup of tea. A friend who frequently travels between the U.S. and China remarked that "*the app wants you to feel that after numerous subconscious interactions and clicks, you have saved a big deal.*" The app design mechanism gives users an illusion of gaming the system or winning against merchants.

(2) **Gaming the Rating System**: Ratings for products and services are ubiquitous in superapps, but I noticed that they are often inflated due to biased sampling. Many restaurants and cafes offer incentives—such as free desserts, soup, or drinks—for customers who leave positive reviews. As a result, most ratings are clustered between 4.5 and 4.9 out of 5, making it difficult to differentiate between options and undermining the reliability of the ratings system. These similar ratings further lead to my decision paralysis in ordering food or finding a haircut. I found myself spending an unintended amount of energy comparing options, influenced by a fear of missing out (*FOMO*), despite not being particularly picky in the U.S..

(3) **Difficulty in Multi-Tasking within Apps**: Since superapps integrate multiple functions with deep information architectures, using different functions simultaneously can be problematic. For example, I was using *Meituan* (美团) to search for a gym membership, navigate with its embedded map to the gym, and rent a bike—all within the same app. Switching between these functions required traversing different branches of the app's hierarchy, while also contending with ads and pop-ups, making the process cumbersome. In the U.S., the different functions are usually provided by various apps, and I can simultaneously keep them open and switch them easily.

These superapps are integrated into every aspect of modern living in China. I felt that my life was intertwined with them—they subtly influenced my perceptions, decisions, and behaviors both online and offline. While I appreciate the convenience they offer, I sometimes feel trapped by the algorithms they've devised. Without these apps, I might not even know what I want or need.

### 5.4 Navigating Chinese Technosocial Practices and Norms: Live Streaming, Payment Systems, and Privacy Trade-offs

Technosocial practices and norms are the socially constructed rules, behaviors, and expectations that emerge around the use and interaction with technology. These norms are shaped by both technological capabilities and social factors, influencing how people behave, perceive, and interact with each other and with technology in various contexts. During my initial ten days in China, I encountered several distinct technosocial norms that contrasted sharply with my experiences in the United States. Adapting to these norms required navigating unfamiliar practices while reconciling them with my existing habits and expectations.

### 5.4.1 Live Stream and Short Video Culture.

Live streaming and short video recording are ubiquitous in China, permeating nearly every public space, including scenic spots, shopping malls, restaurants, parks, and gyms. I observed numerous live streamers setting up tripods or using phone mounts to broadcast their activities. Some streamers were audacious, loudly engaging with their online audiences in random places, while others quietly streamed from restaurants while eating. Unlike in the United States, where I can often avoid being passively live-streamed, in China, the omnipresence of live streamers made it nearly impossible. For instance, at the gym near my stay, free tripods were available for patrons to use, resulting in constant live streaming from various parts of the facility. Consequently, I had to adjust by pretending that being live-streamed did not affect my workout routine.

Additionally, watching live streams and short videos is equally common among the general population. In local stores, waitstaff often watched short videos or live streams during downtimes. While working on my projects in cafes, I noticed that many customers were engrossed in short videos, exhibiting what I term the "phone smirk"—a subtle smile or laugh triggered by the content they were viewing. My own usage of short videos increased significantly compared to my time in the United States, which I attribute to the seamless integration of these features within superapps. Personally, short videos served primarily as a source of entertainment, easily forgotten the next day. Yet, collectively they contributed to a pervasive culture of constant digital engagement.

### 5.4.2 Mobile Payment Systems: Beyond Credit Cards.

In the United States, credit card payments are the predominant mode of transaction, with mobile payment solutions like *Apple Wallet* serving as digital repositories for these cards. In contrast, Chinese mobile payment systems such as *WeChat Pay* (微信支付) and *Alipay* (支付宝) have expanded the concept of mobile payments to encompass a wide array of services. These superapps not only facilitate payments for products and services but also enable money transfers, sending red envelopes, purchasing financial products, paying for public transportation, and even highway tolls. This extensive integration has rendered credit cards nearly obsolete in China, especially among small merchants who rarely accept cash or credit cards [18, 36].

This fundamental difference poses challenges for travelers from the West. For example, a Canadian friend of mine struggled to adapt to the multifunctional payment options within *WeChat Pay* (微信支付) initially, which include features like red envelopes, QR code scanning from merchants, and peer-to-peer transfers. Unlike the straightforward "scan and pay" method associated with credit cards, Chinese mobile payments offer multiple pathways for transactions and transferring, requiring users to understand the subtle differences and corresponding interfaces. It took me considerable time to comprehend these differences, only fully understanding them after conducting research for this study.

### 5.4.3 Trade-Off Between Convenience and Privacy.

In China, phone numbers are intrinsically linked to personal identity through the real-name registration system, which is required for signing up for various services (手机实名制). Unlike many U.S. apps that allow registration via email, Chinese apps typically mandate phone number verification, making phone numbers as ubiquitous as Google sign-ins in the West. Initially, I was reluctant to share my phone number due to the fear of robocalls in the U.S. However, service providers—such as hairdressers, dentists, and cafe waitstaff—persuaded me to provide my number for using coupons and streamline future transactions. This trade-off between convenience and privacy became a recurring theme in my interactions.

For another instance, when signing up for a local gym membership, I was informed that face recognition would be used for entry. Despite my reluctance and requests for alternative methods, the gym insisted on this requirement, promising secure handling of my data. Finally, I caved and signed up for the face recognition. After a while, the convenience of not needing to interact with staff or manage a physical keycard made me not linger on the privacy concerns, especially as I noticed all others treated it as normal. While China is often seen as the most convenient country for getting things done with mobile apps [38], this convenience often comes at the cost of privacy. My experience shows that these trade-offs are unavoidable for adapting to modern life in China.

## 6 Discussion and Reflection

### 6.1 Crossing the Digital Divide: From Outsider to Insider

Throughout my initial ten-day trip to China, I found myself in a constant state of learning—new technology interfaces, new concepts, and new ways of using them. Initially treated as an outsider, I often felt embarrassed and perplexed by the unfamiliar technosocial practices and norms. With the support of my parents, relatives, and numerous strangers who kindly (or sometimes bluntly) offered guidance, I underwent a rapid adaptation. By the end of the ten days, I could confidently navigate everyday tasks without needing to ask for assistance. My willingness to adapt and actively engage with these new practices allowed me to integrate into the local technosocial environment and finally become part of the in-group. I discussed this experience with friends who, like me, travel between the U.S. and China. They echoed my sentiments, describing a "reverse technological culture shock" that typically takes about a week to adjust.

However, not everyone adjusts as readily. Technology can create divides and exacerbate social disparities [13, 17, 37]. Individuals who cannot or choose not to adapt—due to language barriers, systemic obstacles (such as foreigners' lack of local phone numbers for signing up [23]), unfamiliarity, or resistance to change—may find themselves excluded from essential services or social interactions when moving or traveling in different cultures. The ability to adjust and use Chinese mobile technologies quickly divides users into perceived in-groups and out-groups and determines access to services and participation in daily life.

My experience highlights a gap and opportunity, as also pointed out by recent papers [7, 38], in our understanding within the HCI community regarding how cross-culture travelers perceive, experience, and adjust to varying technosocial practices. For example, would children who left China at a young age, with limited reading proficiency in Chinese, struggle to navigate and adapt to these apps upon return? Similarly, would older adults who emigrated long ago and seldom travel back—those in their 50s, 60s, or older—find it even more challenging to comprehend and adjust? Additionally, how do foreigners and expatriates experience using these localized apps? As the world becomes increasingly interconnected, with people frequently crossing cultural and technosocial boundaries, it is crucial to understand how individuals straddle these differences and adjust to respective technosocial practices. I hope my experience can offer some understanding to our HCI community for designing better support for all users navigating cross-cultural technological landscapes.

### 6.2 Smartphone Attachment: A Universal Phenomenon

While specific mobile technosocial practices vary across cultures, there is a universal trend of increasing time spent on smartphones worldwide [11, 35]. Over time, the evolution of technology, propelled by business goals and market competition, has become increasingly adept at capturing our engagement and integrating deeply into every aspect of our lives. I have witnessed this phenomenon in both Chinese society and the United States, as well as around the globe during my travels. Highly successful apps like TikTok (originally *Douyin* "抖音" in China), which first captivated users' attention and time domestically, have demonstrated their ability to replicate this success internationally, notably in the United States. Following current trends in the U.S.—with time spent on such apps more than doubling compared to five years ago [14]—the culture of live streaming and short videos phenomenon I observed in China is likely to grow and garner even more engagement globally.

Despite our diverse cultures, with different preferences for food, clothing, social structures, and more, smartphone usage and the behaviors it induces have become unifying factors experienced across cultures. As an HCI researcher, I find it both fascinating and concerning to observe how quickly technology like smartphones can transform people across cultures within less than two decades. As we enter a new era of artificial intelligence, it is likely that the success of mobile apps will be replicated and even amplified in this domain, dramatically changing our behaviors in even shorter periods. How HCI researchers and practitioners respond to, propel, or mitigate these technology-induced global trends across cultures is a critical question that warrants ponderation.

### 6.3 Variations in App Design Across Cultures

I recall a time when Chinese web and mobile applications closely resembled those in the United States. Initially, many apps in China began by emulating the success of their U.S. counterparts. However, as these apps iterated on their business models and adapted to cultural contexts, they started to compete, evolve, and tailor their offerings to the specific needs of their users and the broader cultural and societal backdrop. The emergence of superapps in China exemplifies this transformation.

There is abundant research that considers cultural factors influencing app design. For example, one study indicates that factors influence the characteristics of app reviews, suggesting that user feedback is deeply rooted in cultural contexts [30]. Another found usability perceptions can vary markedly across cultures, affecting how apps are rated and adopted. What is considered user-friendly in one culture may not hold the same value in another [6]. Additionally, Chopdar et al. found that perceived risks associated with mobile shopping apps differ across countries, highlighting that cultural factors play a crucial role in shaping user intentions and behaviors [9].

These findings, taken together, challenge the Western-centric view of design heuristics widely adopted in UX design since the early 1990s [28]. The limitations of this perspective in understanding people from other cultures have been underscored by research on "WEIRD" (Western, Educated, Industrialized, Rich, and Democratic) populations, which suggests that findings based on Western contexts may not be universally applicable [21]. As this Western-centric approach extends to app design and development, it potentially overlooks the diverse needs and expectations of users in different cultural settings.

A complex interplay of values, usability perceptions, and user expectations causes apps to vary across cultures. These factors cause "*neutral*" technologies to diverge in apps, becoming diverse and adaptive to local contexts. As a traveler, I have witnessed different design styles, interaction methods, and conceptual differences related to apps across various countries. Initially, this often led to frustration, making me wonder why they did not follow the *canonical models* I was accustomed to. However, after my ten-day experience living in China, I began to accept and appreciate the learning process, which provided me with a window into understanding the culture, people, and society on a deeper level.

## 7 Conclusion

Mobile apps are integral to modern life, influencing how we communicate, shop, travel, and interact. Learning and adapting to the local apps can serve as a gateway to understanding and engaging with its culture. In this autoethnography, I described my initial 10-day experience in China, exploring both the challenges and excitement of using unfamiliar apps, revealing the high integration of these apps in modern Chinese living, the complexity of superapps, and my observation and adjustment to Chinese technosocial practices and norms. While some smartphone use behavior is universal, the specific design and functionality of apps vary substantially in response to local contexts, which leads to different user practices and behaviors. By sharing this personal narrative, I hope to inspire researchers to reflect on how mobile apps and technology in general influence people's cross-cultural adaptation in an increasingly interconnected and digitalized world.

### Acknowledgments

I want to thank my parents for supporting my "working vacation" at home and handling all the "maintenance" tasks. In addition, I appreciate the friends and relatives I met during my travels, who contributed valuable raw materials for this paper. I also thank Shuo Yang, for his knowledge and shared experience of traveling back and forth between the U.S. and China, and Alexis Hiniker for her comments and encouragement. Lastly, I am grateful to the strangers whose technology usage I have observed, and I hope our community can design better technology for them to use in the future.

### References

[1] 2017. All You Need to Know About WeChat Mini Programs. https://chozan.co/blog/all-you-need-know-wechat-mini-programs/

[2] 2024. Bilibili. https://www.bilibili.com/ Accessed: 2024-12-01.

[3] 2024. Douyin. https://www.douyin.com/ Accessed: 2024-12-01.

[4] 2024. Figma: The Collaborative Interface Design Tool. https://figma.com

[5] 2024. Wikipedia: Super-app. https://en.wikipedia.org/wiki/Super-app

[6] Lamya Alabdulkarim. 2021. End users' rating of a mHealth app prototype for paediatric speech pathology clinical assessment. *Saudi Journal of Biological Sciences* 28, 8 (2021), 4484–4489.

[7] Jana Bucher, Mila Lazarova, and Jürgen Deller. 2024. Digital technology and global mobility: Narrative review and directions for future research. *International Business Review* (2024), 102294.

[8] Yixin Chen, Yue Fu, Zeya Chen, Jenny Radesky, and Alexis Hiniker. 2024. Extended-Use Designs on Very Large Online Platforms. *arXiv preprint arXiv:2411.12083* (2024).

[9] Prasanta Kr Chopdar, Nikolaos Korfiatis, VJ Sivakumar, and Miltiades D Lytras. 2018. Mobile shopping apps adoption and perceived risks: A cross-country perspective utilizing the Unified Theory of Acceptance and Use of Technology. *Computers in Human Behavior* 86 (2018), 109–128.

[10] James Clifford and George E Marcus. 2023. *Writing culture: The poetics and politics of ethnography.* Univ of California Press.

[11] Deloitte. 2024. More than 7 in 10 use their smartphone as soon as they wake up, Deloitte research finds. https://www.deloitte.com/ie/en/about/press-room/consumer-trends-smartphone-usage.html Accessed: 2024-12-05.

[12] Norman K Denzin and Yvonna S Lincoln. 1996. Handbook of qualitative research. *Journal of Leisure Research* 28, 2 (1996), 132.

[13] Jonathan Donner. 2015. *After access: Inclusion, development, and a more mobile Internet.* MIT press.

[14] Fabio Duarte. 2024. Average Time Spent on TikTok Statistics. https://explodingtopics.com/blog/time-spent-on-tiktok Accessed: 2024-12-05.

[15] Margot Duncan. 2004. Autoethnography: Critical appreciation of an emerging art. *International journal of qualitative methods* 3, 4 (2004), 28–39.

[16] Carolyn Ellis, Tony E Adams, and Arthur P Bochner. 2011. Autoethnography: an overview. *Historical social research/Historische sozialforschung* (2011), 273–290.

[17] Virginia Eubanks. 2012. *Digital dead end: Fighting for social justice in the information age.* MIt Press.

[18] Michelle Evans. 2017. How China Won The Race To Be A Mobile-First Commerce Nation. https://www.forbes.com/sites/michelleevans1/2017/04/12/how-china-won-the-race-to-being-considered-a-mobile-first-commerce-nation/

[19] Clifford Geertz. 2008. Thick description: Toward an interpretive theory of culture. In *The cultural geography reader.* Routledge, 41–51.

[20] Daniel Harrison and Marta E Cecchinato. 2015. " Give me five minutes!" feeling time slip by. In *Adjunct Proceedings of the 2015 ACM International Joint Conference on Pervasive and Ubiquitous Computing and Proceedings of the 2015 ACM International Symposium on Wearable Computers.* 45–48.

[21] Joseph Henrich, Steven J Heine, and Ara Norenzayan. 2010. The weirdest people in the world? *Behavioral and brain sciences* 33, 2-3 (2010), 61–83.

[22] Shangui Hu, Hefu Liu, and Guoyin Wang. 2020. How Does Mobile Devices Usage Contribute to Individual's Creativity in Cross-Cultural Settings?. In *Design, Operation and Evaluation of Mobile Communications: First International Conference, MOBILE 2020, Held as Part of the 22nd HCI International Conference, HCII 2020, Copenhagen, Denmark, July 19–24, 2020, Proceedings 22.* Springer, 23–32.

[23] Yuanyuan Huang, Joaquin Estrader, and Jing Song. 2022. Affinity and foreign users' perception about Chinese mobile apps: An integrated view of affective contagion and value-based perspectives. *Electronic Commerce Research and Applications* 53 (2022), 101157.

[24] Erik M Kormos. 2016. *The usage of smartphone technologies by American expatriate teachers as a communication and cultural assimilation tool.* Indiana University of Pennsylvania.

[25] Andrés Lucero. 2018. Living without a mobile phone: an autoethnography. In *Proceedings of the 2018 Designing Interactive Systems Conference.* 765–776.

[26] Adarsh Nair. 2023. Digital Evolution And How It Differs Across America, China And India. https://www.forbes.com/councils/forbestechcouncil/2023/03/06/digital-evolution-and-how-it-differs-across-america-china-and-india/

[27] Moazzam Naseer. 2012. Digital omnivores, social Media and Social Capital: Expatriates interactions using smartphones in Stockholm.

[28] Jakob Nielsen and Rolf Molich. 1990. Heuristic evaluation of user interfaces. In *Proceedings of the SIGCHI conference on Human factors in computing systems.* 249–256.

[29] OpenAI. 2024. ChatGPT. https://chatgpt.com/

[30] Dennis Pagano and Walid Maalej. 2013. User feedback in the appstore: An empirical study. In *2013 21st IEEE international requirements engineering conference (RE).* IEEE, 125–134.

[31] Alan Peshkin. 1985. Virtuous subjectivity: In the participant-observer's I's. *Exploring clinical methods for social research* 267 (1985), 281.

[32] Gary Pritchard, John Vines, Pam Briggs, Lisa Thomas, and Patrick Olivier. 2014. Digitally driven: how location based services impact the work practices of London bus drivers. In *Proceedings of the SIGCHI Conference on Human Factors in Computing Systems.* 3617–3626.

[33] Amon Rapp. 2018. Autoethnography in human-computer interaction: Theory and practice. *New directions in third wave human-computer interaction: Volume 2-Methodologies* (2018), 25–42.

[34] Ulrike Schultze. 2000. A confessional account of an ethnography about knowledge work. *MIS quarterly* (2000), 3–41.

[35] Backlinko Team. 2024. Smartphone Usage Statistics. https://backlinko.com/smartphone-usage-statistics Accessed: 2024-12-05.

[36] Dominique Torre and Qing Xu. 2020. Digital payments in China: adoption and interactions among applications. *Revue d'économie industrielle* 172 (2020), 55–82.

[37] Alexander JAM Van Deursen and Jan AGM Van Dijk. 2019. The first-level digital divide shifts from inequalities in physical access to inequalities in material access. *New media & society* 21, 2 (2019), 354–375.

[38] Yuling Wang, Martin Lockett, and Abby Jingzi Zhou. 2024. Digitalization and expatriate cross-cultural adjustment: the role of mobile apps. *Journal of Global Mobility: The Home of Expatriate Management Research* (2024).

[39] Xiao Xiao and Hiroshi Ishii. 2016. Inspect, embody, invent: a design framework for music learning and beyond. In *Proceedings of the 2016 CHI Conference on Human Factors in Computing Systems.* 5397–5408.